\documentclass[]{aa}  

\usepackage{natbib}
\bibpunct{(}{)}{;}{a}{}{,}
\usepackage{graphicx}
\usepackage{revsymb}	
\usepackage{amsmath}	
\usepackage[usenames]{color}	
\usepackage{epstopdf}	
\usepackage{txfonts}
\usepackage{amssymb}
\usepackage{color}
\usepackage{geometry} 
\usepackage[parfill]{parskip} 
\usepackage{amssymb}
\usepackage{slantsc}
\usepackage{array}
\usepackage{url}
\usepackage{amsfonts}
\usepackage[colorlinks=true,linkcolor=blue, urlcolor=blue, citecolor=blue]{hyperref}
\usepackage{sidecap}
\usepackage{graphicx}
\usepackage{xcolor}
\usepackage{nicefrac}
\usepackage{subfigure}
\usepackage{epsfig}
\colorlet{rouge}{red!70!darkgray}

\begin{document}

\title{Entropy proxy inversions as tracers of the evolution of physical conditions at the base of the solar convective envelope}
\author{G. Buldgen\inst{1} \and A. Noels\inst{1} \and V.A. Baturin\inst{2} \and J. Christensen-Dalsgaard\inst{3} \and S.V. Ayukov\inst{2} \and A.V. Oreshina\inst{2}\and R. Scuflaire\inst{1}}
\institute{STAR Institute, Université de Liège, Liège, Belgium \and Sternberg Astronomical Institute, Lomonosov Moscow State University, 119234,Moscow, Russia \and Stellar Astrophysics Centre, Department of Physics and Astronomy, Aarhus University, 8000 Aarhus C, Denmark}
\date{July, 2024}
\abstract{The Sun is an important calibrator for the theory of stellar structure and evolution. However, the accuracy of our solar evolution models is tightly linked to the physical ingredients that enter their computations. This include, amongst other, the equation of state, the opacities, the transport of chemicals and the modelling of turbulent convection. Deriving model-independent probes of these ingredients is therefore crucial to further test the quality of these ingredients and potentially reveal their shortcomings using observational data.}
{We aim at providing additional constraints on the thermodynamical properties of the solar plasma at the base of the solar convective zone using a revised helioseismic indicator mimicking the properties of the specific entropy in the envelope.}
{We derive a revised entropy proxy for the solar convective envelope, directly accessible using helioseismic structure inversions. We then use solar evolutionary models with various modifications of input physics to study the properties of proxy of the entropy in the convective envelope.}
{We find that the entropy proxy for the solar convective envelope allows to invalidate adiabatic overshooting as a solution to the solar modelling problem and strongly points towards the need for revised opacities. Our results show that this new indicator is a strong diagnostic of the overall evolution of the thermodynamical conditions at the base of the convective zone.}{The new entropy proxy indicator allows for a more accurate characterisation of the conditions at the base of the solar convective zone. While it already allows to rule out overshooting as a solution to the solar modelling problem, its sensitivity to the shape of the opacity modification and the evolution of the properties at the base of the convective zone makes it a powerful helioseismic diagnostic for solar models.}
\titlerunning{Entropy proxy inversions in the solar convective envelope}
\maketitle
\section{Introduction}

The Sun is a crucial laboratory of stellar physics at microscopic and macroscopic scales and a central reference point for stellar evolution \citep[see][and refs therein]{JCD2021}. Thanks to helioseismic and spectroscopic data, supplemented by the detections of neutrinos, solar modellers have accurate and precise constraints on the internal structure and dynamics of our star \citep[see e.g.][ and refs therein]{JCD91Conv,BasuYSun, JCD1996, Basu97BCZ,SchouRota,BasuSun,OrebiGann2021,Appel2022}. A crucial point in the solar structure is the base of the convective zone (BCZ), which is the seat of a variety of complex physical phenomena \citep[see][for a book on the topic]{Hughes2007} and other works \citep[e.g.][and refs therein]{Spiegel1992,Garaud2002,Garaud2008,Acevedo2013,Strugarek2023}. Helioseismology allowed to locate very precisely the radial position where this transition occurs, at $0.713\pm0.001$ solar radii \citep{JCD91Conv, Basu97BCZ}. Additional studies have been devoted to study aspects of the thermodynamical properties of the convective envelope itself \citep{Vorontsov13,VorontsovSolarEnv2014} and its link to the equation of state \citep[see e.g.][]{JCD1991Env,JCD1992,Vorontsov1992,Baturin1995,Elliott1996}. In this work, we discuss the properties of the entropy of the convective envelope. As convection is essentially adiabatic in the deep envelope layers, a local measure of the specific entropy should show that it behaves as a constant. This value, as shown by \citet{Baturin1995}, is a function of the chemical and thermal stratification at the BCZ. In this work, we build on our previous study \citep{BuldgenS} and present a new entropy proxy and study in detail its properties and link with convective penetration and opacity at the BCZ.

We start by presenting and discussing the equations to derive a new entropy proxy for the solar envelope in Sect. \ref{Sec:SPlateau}, and an inversion of this proxy as a function of normalized radius in the Sun is also presented. The physical information carried by this entropy proxy is then studied in detail in solar calibrated models in Sect. \ref{Sec:BCZ}, where we investigate the effects of adiabatic overshooting at the BCZ and an opacity increase close to the BCZ on the properties of the entropy proxy in the convective envelope. We then study the dominant factors influencing the height of the plateau in the convective zone in Sect. \ref{Sec:DominantFact}, showing the potential of this quantity to help better constrain the thermodynamical conditions at the BCZ. We also discuss in Sect. \ref{Sec:ThermoQuantities} some observed changes in key thermodynamical quantities such as the first adiabatic exponent, the specific heat at constant volume and the electronic density between the two families of models and discuss potential future studies and our conclusions in Sect. \ref{Sec:Conc}.

\section{Entropy proxy in the solar envelope}\label{Sec:SPlateau}

The stratification in the deep convective envelope of the Sun is almost adiabatic, which motivated \citet{BuldgenS} to define a proxy for the entropy, denoted $S_{5/3}=P/\rho^{5/3}$ in the convective envelope which was used both for helioseismic and asteroseismic inversions \citep{BuldgenS,Buldgen2018,Betrisey2022}. The equations to derive the kernels for this variable are provided in \citet{Buldgen2017Ker} and a discussion of the behaviour of this physical quantity is also given in \citet{Buldgen2018,Betrisey2022}. 

\begin{figure}
	\centering
		\includegraphics[width=8cm]{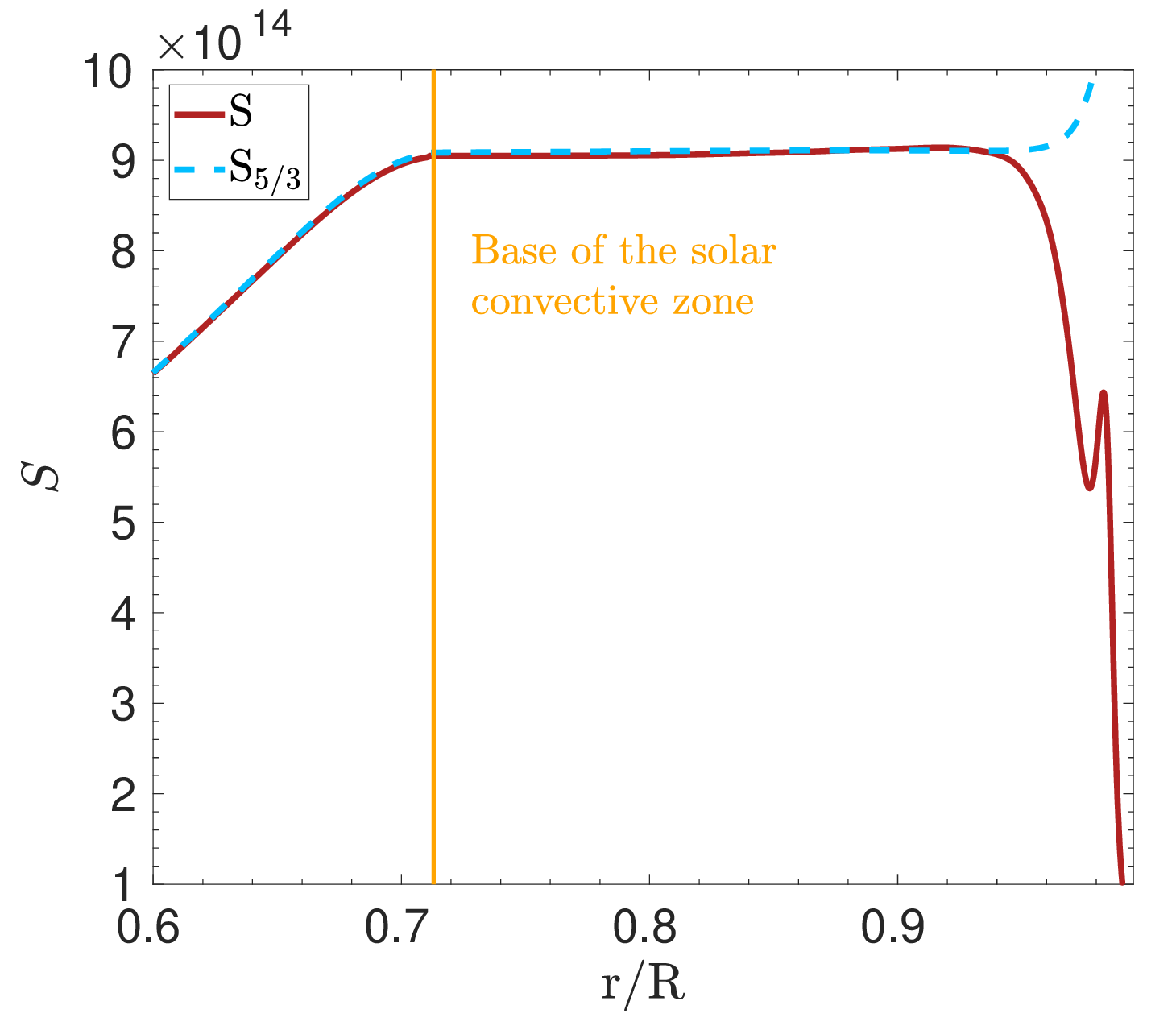}
         \caption{Comparison between the two entropy proxies, $S$ and $\rm{S_{5/3}}$, showing the divergence of the proxy $\rm{S_{5/3}}$ at the surface and the plateau-like behaviour of both proxies in the deep convective layers.}
		\label{Fig:SvsS53}
\end{figure}

One of the key issues of the $S_{5/3}$ variable is that it takes extreme values at the surface. While this can be mitigated and the inversions remain robust, it is clear that damping these effects is desirable, while keeping the main property of the indicator, namely the plateau-like behaviour in the convective zone. This is done by defining a new variable, $S=P/\rho^{\Gamma_{1}}$, which is essentially a refinement of the previous indicator with $\Gamma_{1}$ the first adiabatic exponent $\Gamma_{1}=\frac{\partial \ln P}{\partial \ln \rho}\Big|_{\mathcal{S}}$, $\mathcal{S}$ the entropy. A comparison between the behaviour of $S$ and $S_{5/3}$ is illustrated in Fig. \ref{Fig:SvsS53}. Two striking differences can be noted in this profile. First, the amplitude of the variable in the higher convective envelope is much lower. Second, due to the direct use of $\Gamma_{1}$ in the expression, a direct trace of helium ionization in the upper envelope is visible in the profile. These behaviours make this revised indicator far more adapted to probe the solar convective envelope, while the plateau-like behaviour has been fully kept. 

If we express the variations of entropy using thermodynamic variables, we find that
\begin{align}
d \mathcal{S}=\left. \frac{\partial \mathcal{S}}{ \partial \ln P} \right\vert_{\rho} d \ln P + \left. \frac{\partial \mathcal{S}}{ \partial \ln \rho} \right\vert_{P} d \ln \rho,\label{eq:EntroDiff}
\end{align}
where we use the following relations
\begin{align}
\left. \frac{\partial \mathcal{S}}{ \partial \ln P} \right\vert_{\rho} &= \frac{C_{V}}{\chi_{T}}, \\
\left. \frac{\partial \mathcal{S}}{ \partial \ln \rho} \right\vert_{P} &= -\frac{C_{P} \chi_{\rho}}{\chi_{T}}
\end{align}
with $\chi_{T}=\left. \frac{\partial \ln P}{ \partial \ln T} \right\vert_{\rho}$, $\chi_{\rho}=\left. \frac{\partial \ln P}{ \partial \ln \rho} \right\vert_{T}$, $C_{V}=\left.\frac{\partial \mathcal{S}}{\partial \ln T} \right\vert_{V}$ and $C_{P}=\left.\frac{\partial \mathcal{S}}{\partial \ln T} \right\vert_{P}$.
This allows to rewrite Eq. \ref{eq:EntroDiff} as
\begin{align}
d \mathcal{S}=\frac{C_{V}}{\chi_{T}} \left(d \ln P - \Gamma_{1} d \ln \rho \right), \label{eq:EntroDiff2}
\end{align}
where we used that the first adiabatic exponent is also equal to $\Gamma_{1}=\frac{C_{P} \chi_{\rho}}{C_{V}}$. The ideal gas approximation provides $\chi_{T}=\chi_{\rho}=1$ and $\gamma=\frac{C_{P}}{C_{V}}$. In the case where $\gamma$ (or $\Gamma_{1}$) is constant, the solution to an adiabatic change, meaning $\int d\mathcal{S}=0$, is the analytical relation
\begin{align}
\frac{P}{\rho^{\gamma}}= constant,
\end{align}
which links our entropy proxy, $S$, to the thermodynamical entropy $\mathcal{S}$ in the regions where $\Gamma_{1}$ is a constant. The relation is, however, not valid where $\Gamma_{1}$ shows strong variations, such as the helium and hydrogen ionization zones.

The equations to derive the kernels for $(S,\Gamma_{1})$ structural pair are very similar to the ones for the $(S_{5/3}, \Gamma_{1})$ pair and can directly be obtained following the approach of \citet{Buldgen2017Ker}. They are provided below

\begin{align}
&y \Gamma_{1}\frac{d^{2}\mathcal{K}}{dy^{2}} - \left[\left( \frac{m \rho}{2y^{1/2}P} - 2y \frac{d \Gamma_{1}}{dy}- \left( 3 - \frac{2 \pi \rho y^{3/2}}{m} \right) \right)\Gamma_{1} \right]  \frac{d\mathcal{K}}{dy} - \nonumber \\
& \left[ \frac{m}{4y^{1/2}P} \frac{d}{dx}\left( \frac{\rho}{y^{1/2}}  \right) - \frac{1}{4}\frac{d^{2}\Gamma_{1}}{dy^{2}} + \left( \frac{\rho m}{2y P} \right)^{2} - \left( \frac{5}{2} - \frac{2\pi \rho y^{3/2}}{m}\right) \frac{d \Gamma_{1}}{dy}\right] \mathcal{K} \nonumber \\
& =-y \frac{d^{2}\mathcal{K^{'}}}{dy^{2}} - \left( 3 -\frac{2 \pi \rho y^{3/2 }}{m} \right) \frac{d\mathcal{K^{'}}}{dy}, \nonumber
\end{align}

where $\mathcal{K^{'}}=K^{n,\ell}_{\rho,\Gamma_{1}}$, $\mathcal{K}=K^{n,\ell}_{S,\Gamma_{1}}$, and $x=r/R$, $y=x^{2}$. The second kernel of the pair is obtained from the simple relation
\begin{align}
K^{n,\ell}_{\Gamma_{1},S}=K^{n,\ell}_{\Gamma_{1},\rho}+\Gamma_{1}\ln \rho K^{n,\ell}_{S,\Gamma_{1}},
\end{align}

The behaviour of the kernels of the $(S,\Gamma_{1})$ pair is extremely similar to that of the $(S_{5/3}, \Gamma_{1})$ pair, as a result of the intrinsic similarities between $S_{5/3}$ and $S$.

An illustration of the inversion for the profile of $S$ in the solar interior for a reference model built with the Grevesse and Noels abundance \citet[GN93]{GrevNoels} (similarly to Model~S in \citet{JCD1996}) and a recent standard solar model built with the \citet[AAG21]{Asplund2021} abundances is provided in Fig. \ref{Fig:SInv}. We also provide in Appendix \ref{sec:AppendixA} an example of inversion using artificial data, where we can see that the height of the plateau determined from the inversion is in excellent agreement with the actual model differences. The overall behaviour is very similar to that of $S_{5/3}$, as illustrated in \citet{BuldgenS}.

\begin{figure}
	\centering
		\includegraphics[width=9cm]{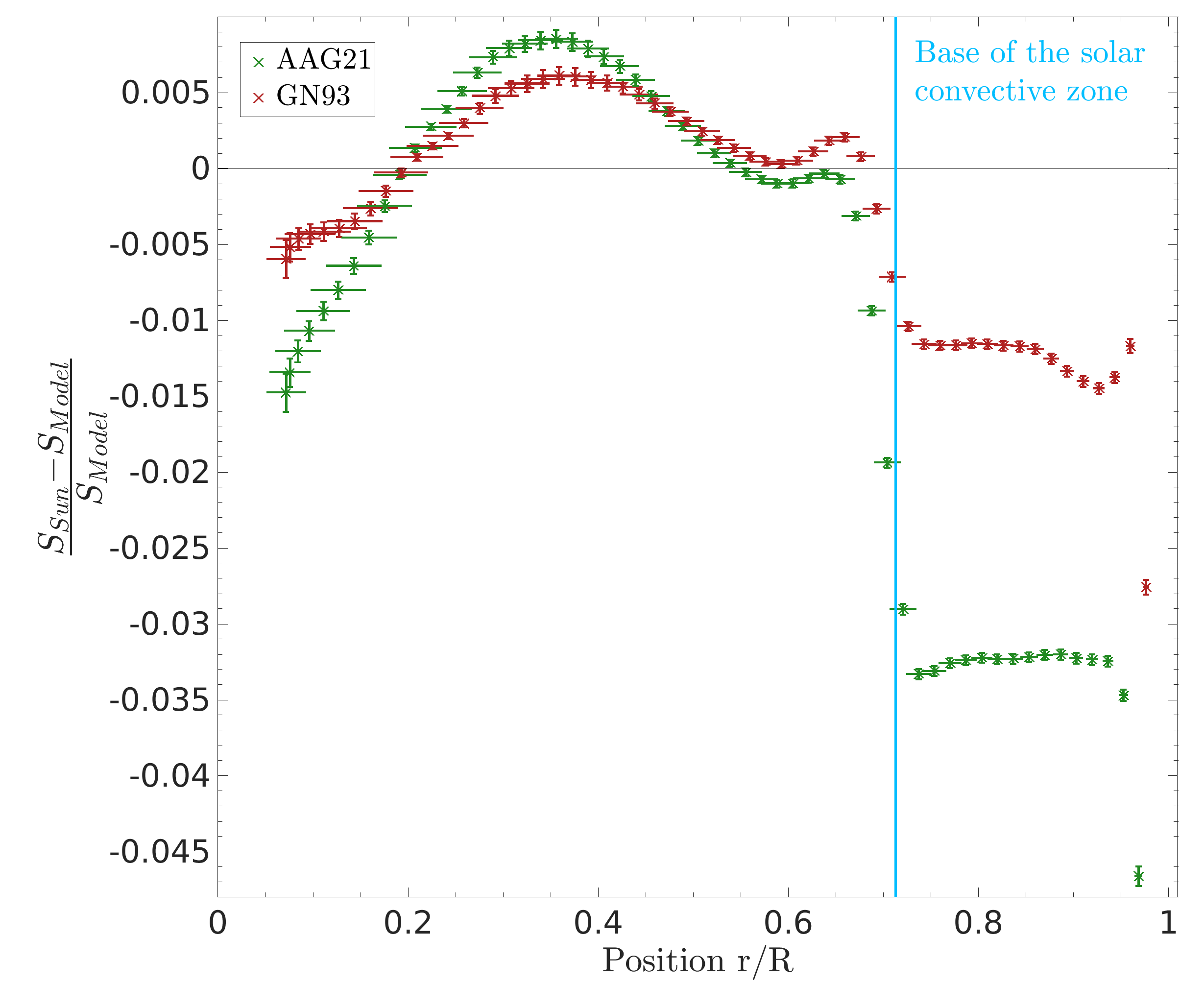}
         \caption{Inversion results for the entropy proxy profile, S, as a function of normalized radius for a standard solar model including the current reference photospheric abundances and the ones used in Model S from \citet{JCD1996}.}
		\label{Fig:SInv}
\end{figure} 

\begin{figure}
	\centering
		\includegraphics[width=8cm]{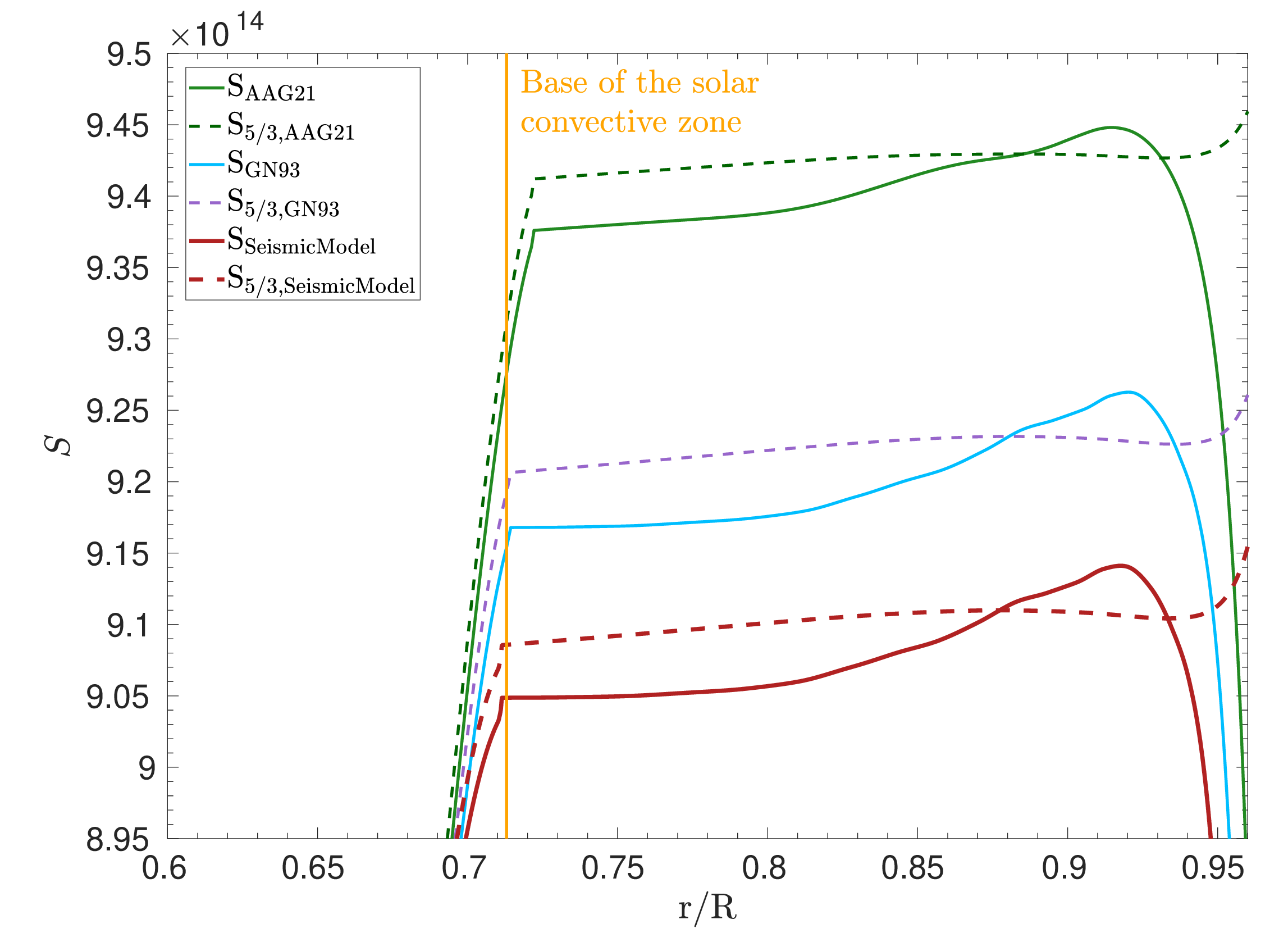}
         \caption{Close comparison between the entropy proxies $S$ and $S_{5/3}$ in the deep convective layers for models including various solar abundances. The same profiles for a seismic model are also provided for comparison.}
		\label{Fig:SModels}
\end{figure}

We also illustrate in Fig. \ref{Fig:SModels} the behaviour of the profile of the entropy proxy itself for both standard solar models and one from a seismic model of \citet{Buldgen2020}. This plot illustrates that the observed trend for the entropy proxy is indeed very similar to that observed in \citet{BuldgenS}, namely the entropy proxy plateau is too high in standard solar models. This is reminiscent of what is observed in density inversions, altough the flat profile observed here is directly due to the physical properties of the convective envelope. In what follows, we will discuss the required changes to solar models that may allow to lower the height of the entropy plateau, following the initial observations made in \citet{Buldgen2019}. A first conclusion that can be drawn is that the entropy proxy plateau of models built with the GN93 abundances that were used for Model~S in \citet{JCD1996} still have a significantly too high plateau. A similar issue with high-metallicity models was already observed in \citet{BuldgenS} and \citet{Buldgen2019}. 

In what follows, we will use solar calibrated models computed with the Liège Stellar Evolution Code \citep[CLES][]{ScuflaireCles}, using as standard physical ingredients the AAG21 abundances \citep{Asplund2021}, the SAHA-S equation of state \citep{Gryaznov04,Gryaznov06,Baturin2017}, the OPAL opacity tables \citep{OPAL}, NACRE II nuclear reaction rates \citep{Xu2013} and the Model-C atmosphere from \citet{Vernazza}. The mixing of chemical elements is treated following \citet{Thoul} and using the screening coefficients of \citet{Paquette}, including the effects of partial ionization. The calibrations are computed using three free parameters: the initial hydrogen and heavy element mass fraction, $X_{0}$ and $Z_{0}$, as well as the mixing length parameter of convection, $\alpha_{\rm{MLT}}$ while the current solar luminosity, radius and metallicity are used as constraints. 
 
\section{Influence of the BCZ conditions on the position of the plateau}\label{Sec:BCZ}

\begin{figure}
	\centering
		\includegraphics[width=8.5cm]{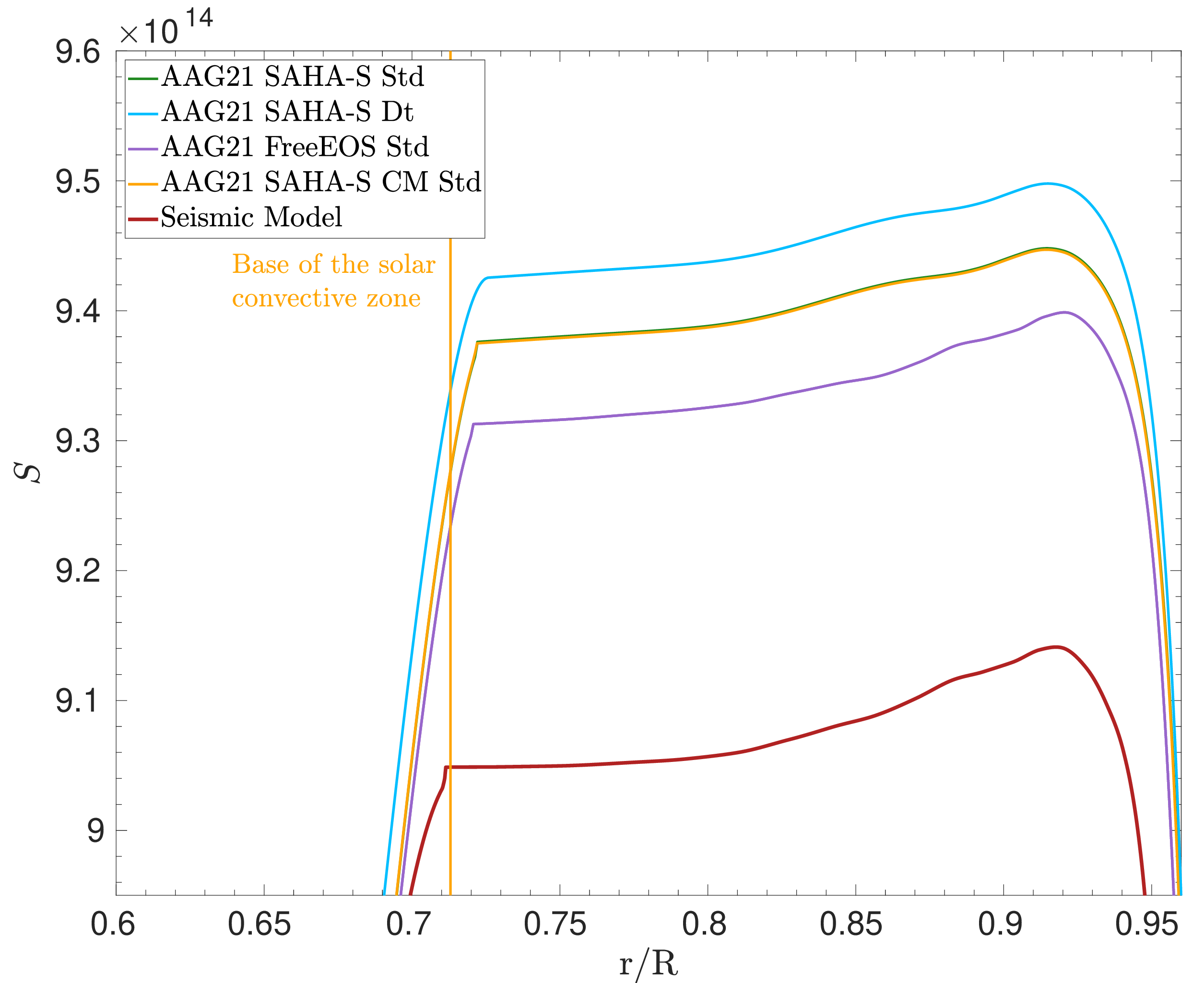}
         \caption{Entropy proxy plateau as a function of normalized radius at the age of the Sun for solar models including the following effects: standard model with reference physics, denoted ``SAHA-S Std'', inclusion of macroscopic mixing at the BCZ, denoted ``SAHA-S Dt'', following \citep{Buldgen2025Be}, change of the equation of state ``FreeEOS-Std'', change of the modelling of convection in the envelope, denoted ``SAHA-S CM Std''. The entropy proxy profile of a seismic model is provided for comparison.}
		\label{Fig:SModMix}
\end{figure}

The height of the entropy proxy plateau seems to be highly dependent on the BCZ conditions. As shown in Fig. \ref{Fig:SModels}, a change in reference abundances has a direct impact on its position. Similarly, the effects of macroscopic mixing at the BCZ also have a direct signature on the height of the plateau, as a result of their impact on the temperature gradients at the BCZ. This effect is illustrated in Fig. \ref{Fig:SModMix}. A direct effect of a change in opacity was also observed in \citet{BuldgenS}, where the use of the OPLIB opacity tables \citep{Colgan} showed a significant improvement in the height of the $S_{5/3}$ plateau. 

While the definition of the new entropy proxy $S$ includes an explicit dependency in $\Gamma_{1}$, the observed differences between the Sun and solar models cannot be explained by variations in the equation of state alone. This is also illustrated in Fig. \ref{Fig:SModMix}, where we plot the profile of $S$ for calibrated solar models computed with the FreeEOS \citep{Irwin} and the SAHA-S equation of state. Significant differences are observed, but still far from what is required to reproduce the seismic value of the entropy proxy. The differences between the two models that have been built using the same reference opacities, abundances and mixing scheme may, however, indicate a strong dependency of the entropy proxy on other thermodynamical variables that differ between the FreeEOS and SAHA-S equations of state. As such, further detailed comparisons between equations of state should still be considered of high priority for both solar modelling and opacity computations \citep{Pradhan2024EOS}. Another ingredient of solar models that may influence the height of the entropy proxy plateau is the modelling of convection itself. To assess its impact, we calibrated a solar model using the Full Spectrum of Turbulence depiction of convection from \citet{Canuto91}. From Fig. \ref{Fig:SModMix}, we can see that the variation of the entropy plateau is negligible.

The importance of the entropy of the solar convective envelope has been discussed in previous publications \citep{BuldgenS,Buldgen2019}, linking it to the solar modelling problem, and earlier works took a deep interest in the entropy of the convective envelope \citep{Baturin1995}. Indeed, it is well known that solar models using the AAG21 abundances still have issues in reproducing the position of the BCZ. In addition, our works have shown that despite being favoured by helioseismic inversions of the chemical composition of the envelope \citep{BuldgenZ,Buldgen2024Z,Baturin2024}, these models cannot naturally reproduce the height of the entropy proxy plateau. In fact, as shown in Fig. \ref{Fig:SModMix}, if macroscopic mixing is included to actually reproduce the inferred chemical composition of the solar envelope, the situation gets significantly worse, both for the BCZ position and the entropy proxy plateau. 

Two solutions have been invoked in the litterature to replace the BCZ position at the helioseismically inferred location. First, the effects of convective overshooting at the BCZ have been invoked, as one would not expect the mixing-length theory to naturally reproduce the BCZ position of the Sun without any additional mixing \citep[This has been discussed for example in][]{Monteiro94,Zhang2019,Baraffe2022}. Second, another approach to steepen the temperature gradients on the radiative side of the BCZ is to increase opacity, which has been done in a wide range of ad-hoc manners in the litterature \citep[e.g.][]{JCD2009,Ayukov2017,Buldgen2019,Kunitomo2021} and has recently gained more traction thanks to experimental measurements in solar conditions that have revealed disagreements with theoretical computations. 

In the following sections, we investigate the behaviour of the entropy proxy plateau under the effects of such modifications and its link with the thermodynamical properties of the BCZ. All models under study use the SAHA-S equation of state, the OPAL opacities and the AAG21 abundances.

\subsection{Impact of overshooting}\label{Sec:Overshooting}

We start by discussing the effects of adiabatic overshooting at the BCZ. From a helioseismic point of view, it is well known that the temperature gradient transition from the adiabatic gradient to the radiative gradient is located at $0.713\pm0.001$ solar radii. The modelling of the temperature gradient below this limit has been attempted in numerous previous studies \citep{Monteiro94,Rempel04,JCDOV,Zhang2019}, the latest of which using constraints from hydrodynamical simulations. Here, we simply consider the impact of adiabatic overshooting with instantaneous mixing to study the resulting variations of the entropy proxy plateau, irrespective of the actual agreement with helioseismic constraints. We computed a series of calibrated solar models, following a simple standard solar model approach and including increasing values of adiabatic overshooting at the BCZ, $\alpha_{\rm{Ov}}$, from $0.1\rm{H_{P}}$ to $0.65\rm{H_{P}}$, with ${\rm H_{P}}=-dr/d \ln P$ the local pressure scale height. The properties of these models are summarized in Table \ref{tabModelsOv}, where we provide the position, mass coordinate, temperature, density, metallicity (Z) and hydrogen mass fraction (X) at the BCZ. The position provided in Table \ref{tabModelsOv} is that of the transition of the temperature gradients and thus includes the overshooting zone. From a helioseismic point of view, the feature observed in the data is that of the temperature gradient transition, regardless of the fact that it corresponds to the limit provided directly by the mixing length theory or if the mixing region is extended as done in these models.

\begin{table*}[h]
\caption{Global parameters of the solar evolutionary models including adiabatic overshooting at the BCZ.}
\label{tabModelsOv}
  \centering
\begin{tabular}{r | c | c | c | c | c | c}
\hline \hline
\textbf{Name}&\textbf{$\left(r/R\right)_{\rm{BCZ}}$}&\textbf{$\left( m/M \right)_{\rm{BCZ}}$}&\textbf{$\log (\mathit{T}_{\rm{BCZ}})$} & $\mathit{\rho}_{\rm{BCZ}}$ ($\rm{g/cm^{3}}$) & $\mathit{Z}_{\rm{BCZ}}$& $\mathit{X}_{\rm{BCZ}}$\\ \hline
Model $0.1\rm{H_{P}}$&$0.7144$&$0.9770$& $6.335$ & $0.1785$& $0.0139$ & $0.7456$\\
Model $0.15\rm{H_{P}}$&$0.7104$&$0.9761$& $6.343$ & $0.1840$ & $0.0139$ & $0.7453$\\ 
Model $0.2\rm{H_{P}}$&$0.7063$&$0.9752$& $6.352$ & $0.1898$ & $0.0139$ & $0.7450$\\ 
Model $0.25\rm{H_{P}}$&$0.7022$&$0.9744$& $6.361$ & $0.1960$ & $0.0139$ & $0.7450$\\
Model $0.3\rm{H_{P}}$&$0.6980$&$0.9734$& $6.370$ & $0.2020$& $0.0139$ & $0.7444$\\
Model $0.35\rm{H_{P}}$&$0.6936$&$0.9724$& $6.378$ & $0.2100$ & $0.0139$ & $0.7442$\\
Model $0.4\rm{H_{P}}$&$0.6893$&$0.9714$& $6.387$ & $0.2160$& $0.0138$ & $0.7440$\\
Model $0.45\rm{H_{P}}$&$0.6851$&$0.9703$& $6.396$ & $0.2241$ & $0.0138$ & $0.7432$\\
Model $0.5\rm{H_{P}}$&$0.6807$&$0.9691$& $6.405$ & $0.2332$& $0.0138$ & $0.7427$\\
Model $0.6\rm{H_{P}}$&$0.6712$&$0.9665$& $6.423$ & $0.2511$& $0.0138$ & $0.7425$\\
Model $0.65\rm{H_{P}}$&$0.6666$&$0.9651$& $6.432$ & $0.2613$ & $0.0138$ & $0.7421$\\
\hline
\end{tabular}
\end{table*}

\begin{figure}
	\centering
		\includegraphics[width=8cm]{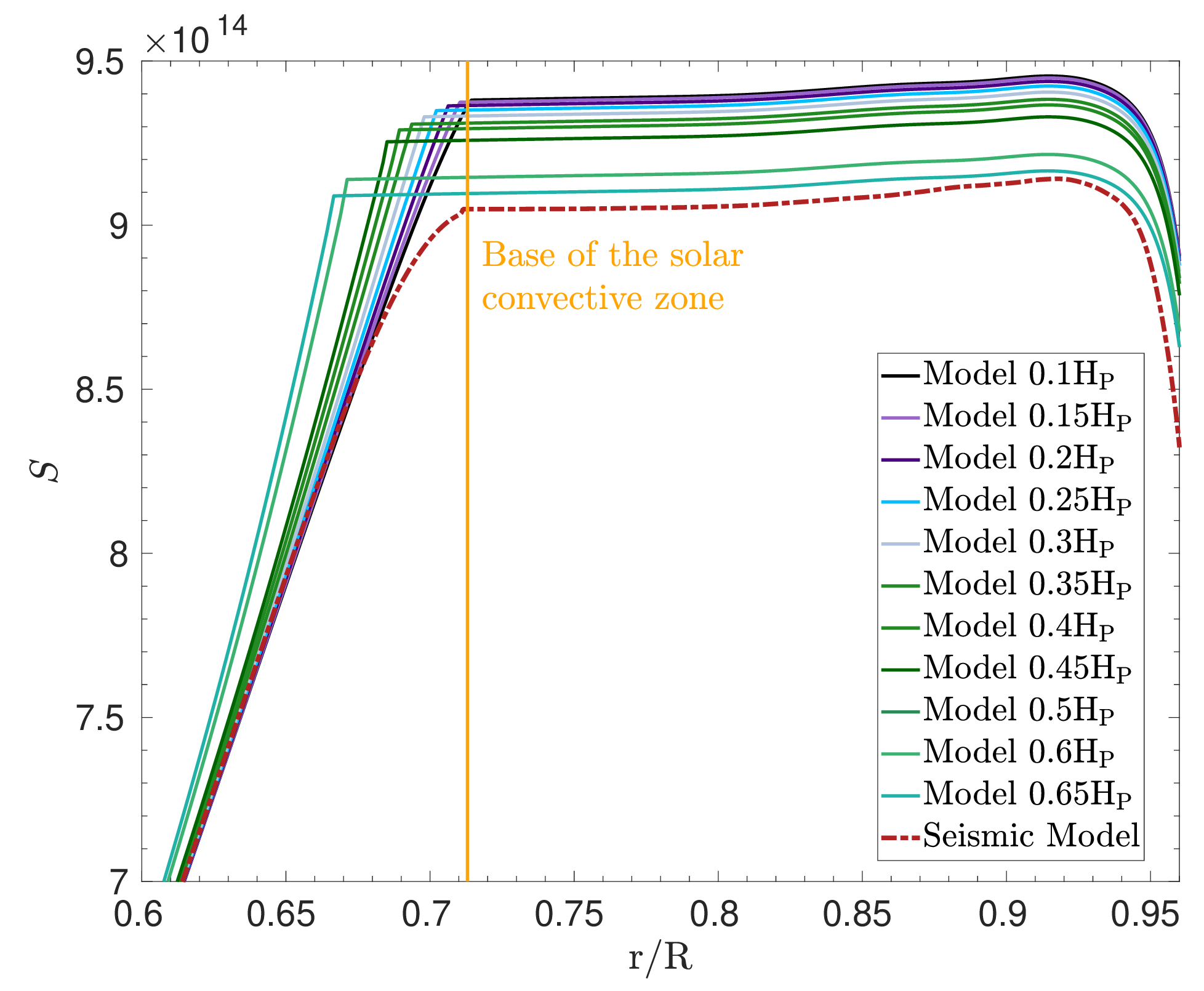}
	\caption{The entropy proxy profile as a function of normalized radius at the age of the Sun for the models of Table \ref{tabModelsOv}. The entropy proxy profile of a seismic model is also provided for comparison.}
		\label{Fig:SOver}
\end{figure}

As expected, increasing the amount of adiabatic overshooting leads to more massive, deeper convective envelope with a higher temperature at the bottom. We can see that the effect on the metallicity of the calibrated solar model is minimal, with a reduction of only $0.0001$ in mass fraction between a model replace at the helioseismic position and one pushing it down to $0.67$ solar radii. The change in both density an temperature is substantial, of about $40 \%$ for the former and $25 \%$ for the latter. Despite these large increases, in stark disagreement with helioseismic constraints, we can see in Fig. \ref{Fig:SOver} that the height of the plateau is still not at the helioseismically determined value, indicated by the seismic model. The behaviour of the plateau is also relatively regular with respect to overshooting, with a deeper convective envelope systematically leading to a lower plateau.

\begin{figure*}
	\centering
		\includegraphics[width=16cm]{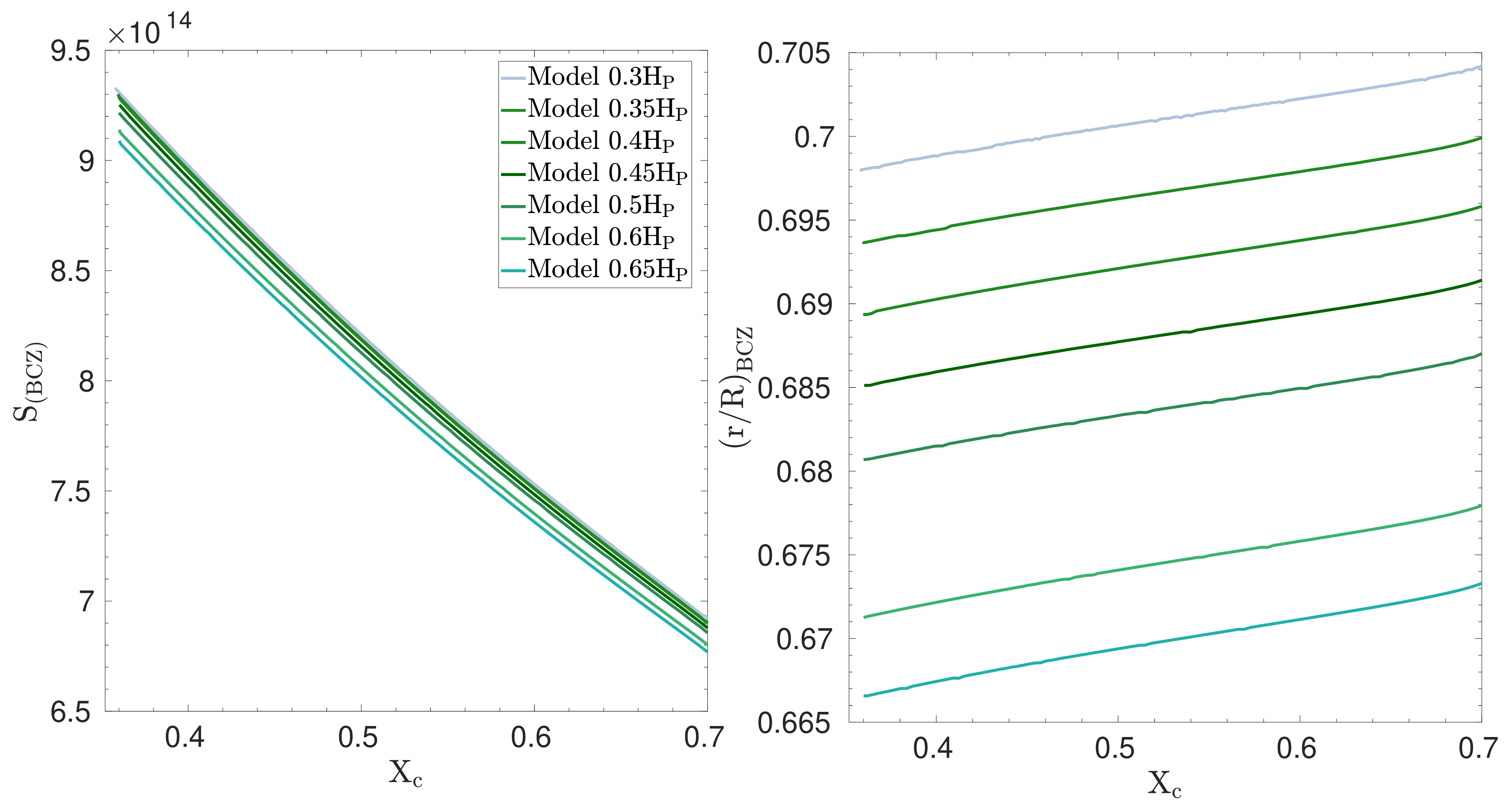}
	\caption{Evolution of the height of the entropy proxy plateau at the BCZ and of the depth of the BCZ as a function of central hydrogen mass fraction for the models of Table \ref{tabModelsOv}. Left panel: evolution of the height of the entropy proxy plateau, S(cz). Right panel: evolution of the position of the BCZ.}
		\label{Fig:SOverEvol}
\end{figure*}

This demonstrates that overshooting is not efficient enough at decreasing the entropy proxy value in the convective envelope and cannot be invoked as the sole solution to the existing discrepancies in solar models. This is further confirmed by recent modelling efforts based on hydrodynamical simulation \citep{Baraffe2022}, which limit the region affected by convective elements to a very narrow layer. This does not mean that overshooting is not acting in solar models, but rather that its effect cannot be invoked to explain the height of the entropy proxy plateau. 
We illustrate in Fig. \ref{Fig:SOverEvol} the evolution during the main-sequence of the position of the base of the convective envelope and of the height of the entropy proxy plateau at the BCZ during the evolution of our solar models with $\alpha_{\rm{Ov}}>0.3\rm{H_{P}}$. As can be seen the behaviour is extremely regular, the BCZ position as a function of time is simply pushed deeper by the overshooting parameter and the height of the plateau is thus lowered throughout the evolution on the main-sequence. We can also see that while the position of the BCZ deepens during the main-sequence, the entropy plateau rises over time indicating that there is indeed an independent constraint brought by the entropy proxy from the BCZ position itself. Indeed, the behaviour of the entropy proxy seems more in line with the evolution of the mass coordinate of the BCZ that increases with time as the convective envelope becomes deeper during the main-sequence, but the temperature at the BCZ decreases. We shall, however, see in the Sect. \ref{Sec:Opacity} that the entropy proxy still provides additional insight to the mass coordinate of the BCZ, meaning that it may serve as an additional diagnostic of the properties of the solar convective envelope. 

\subsection{Impact of an opacity increase}\label{Sec:Opacity}

The other way to steepen temperature gradients at the BCZ and affecting the entropy proxy plateau in the envelope is by increasing the opacity on the radiative side of the border. This was already observed in \citet{Buldgen2019}, while the importance of opacity was already seen in \citet{BuldgenS}. However, these studies did not investigate in detail what was the underlying physical mechanism leading to the variations of the height of the entropy proxy plateau. For example, while it was clearly seen that the amplitude of the opacity increase had an impact on the entropy proxy, it is also clear that the exact functional form of the increase and its dependency with evolution will have a clear impact on the final properties of the solar models. 

To draw a more detailed picture of the influence of the opacity profile on the height of the entropy plateau while keeping a limited number of free parameters, we describe the opacity increase using a simple Gaussian formula, localized around a single temperature as in previous works \citep[e.g.][]{Ayukov2017,Buldgen2019,Kunitomo2021}. This increase is motivated by experimental results \citep{Bailey} indicating a significant discrepancy in iron opacity at BCZ conditions, it is however by no means an accurate depiction of the physical explanation behind the experimental and theoretical opacities but will help us understand the behaviour of the entropy proxy plateau and what parameters influence its position at the solar age. 
Similarly, we note that systematic differences have also been observed for nickel and chromium \citep{Nagayama2019}, although at a much more moderate level and that recent work points towards issues with oxygen \citep{Mayes2025}.

The opacity increase is thus simply described as
\begin{align}
\kappa^{'}=\kappa (1 + \delta),
\end{align}
with $\kappa^{'}$ the modified opacity, $\kappa$ the reference opacity from the table used in the model and $\delta$ the parametrized increase
\begin{align}
\delta=0.01 A \exp^{-\sigma(\log T - \log T_{\rm{ref}})^{2}},
\end{align}
with $A$, $\log T_{\rm{ref}}$ and $\sigma$ being free parameters in our framework. $A=x$ implying an $x\%$ increase of the mean Rosseland opacity derived from the tables. We will therefore test various values for these parameters and see which ones impact the height of the entropy proxy plateau and the thermodynamical conditions at the BCZ. We use the following naming conventions for the models including this opacity increase: Model $A-\sigma-\log T_{\rm{ref}}$. We present out test cases in Table \ref{tabModelsOPAC}, where we considered relatively small variations of $1\%$ in amplitude of the peak, $100$ in $\sigma$ and $0.01$ in $\log T$.

All these models also include a calibrated macroscopic mixing coefficient parametrized as a power law of density as in \citet{Proffitt1991}. They were computed following the approach of \citet{Buldgen2025Be} so that they reproduce the observed lithium and beryllium depletion at the age of the Sun \citep{Wang2021,Amarsi2024}. The effects of macroscopic mixing can be seen from their significantly lower hydrogen mass fraction in the convective envelope in Table \ref{tabModelsOPAC} compared to the values found in the models that only include overshooting in Table \ref{tabModelsOv}.

\begin{table*}[h]
\caption{Global parameters of the solar evolutionary models including an ad-hoc opacity increase at the BCZ.}
\label{tabModelsOPAC}
  \centering
\begin{tabular}{r | c | c | c | c | c | c }
\hline \hline
\textbf{Name}&\textbf{$\left(r/R\right)_{\rm{BCZ}}$}&\textbf{$\left( m/M \right)_{\rm{BCZ}}$}&\textbf{$\log (\mathit{T}_{\rm{BCZ}})$}& $\mathit{\rho}_{\rm{BCZ}}$ ($\rm{g/cm^{3}}$) & $\mathit{Z}_{\rm{BCZ}}$& $\mathit{X}_{\rm{BCZ}}$\\ \hline
Model $14-480-6.35$&$0.7138$&$0.9762$& $6.340$ & $0.1845$& $0.0138$ & $0.7359$\\
Model $14-380-6.35$&$0.7136$&$0.9761$& $6.340$ & $0.1845$ & $0.0138$ & $0.7359$\\ 
Model $14-280-6.35$&$0.7134$&$0.9760$& $6.341$ & $0.1855$ & $0.0138$ & $0.7359$\\ 
Model $14-180-6.35$&$0.7131$&$0.9759$& $6.341$ & $0.1863$ & $0.0138$ & $0.7358$\\
Model $14-80-6.35$&$0.7124$&$0.9756$& $6.342$ & $0.1880$& $0.0138$ & $0.7356$\\
Model $15-280-6.35$&$0.7125$&$0.9758$& $6.342$ & $0.1871$ & $0.0138$ & $0.7357$\\
Model $15-280-6.34$&$0.7126$&$0.9758$& $6.342$ & $0.1867$& $0.0138$ & $0.7357$\\
Model $15-180-6.36$&$0.7124$&$0.9757$& $6.343$ & $0.1881$ & $0.0138$ & $0.7356$\\
Model $15-180-6.35$&$0.7122$&$0.9757$& $6.343$ & $0.1880$& $0.0138$ & $0.7356$\\
Model $15-180-6.34$&$0.7123$&$0.9757$& $6.343$ & $0.1875$& $0.0138$ & $0.7357$\\
Model $15-35-6.36$&$0.7101$&$0.9747$& $6.348$ & $0.1943$ & $0.0137$ & $0.7347$\\
Model $15-80-6.36$&$0.7110$&$0.9751$& $6.345$ & $0.1916$ & $0.0138$ & $0.7354$\\
Model $15-80-6.35$&$0.7115$&$0.9753$& $6.345$ & $0.1903$ & $0.0138$ & $0.7354$\\
Model $15-80-6.34$&$0.7117$&$0.9754$& $6.344$ & $0.1896$ & $0.0138$ & $0.7355$\\
Model $16-180-6.36$&$0.7115$&$0.9754$& $6.345$ & $0.1898$ & $0.0138$ & $0.7356$\\
\hline
\end{tabular}
\end{table*}

\begin{figure}
	\centering
		\includegraphics[width=9cm]{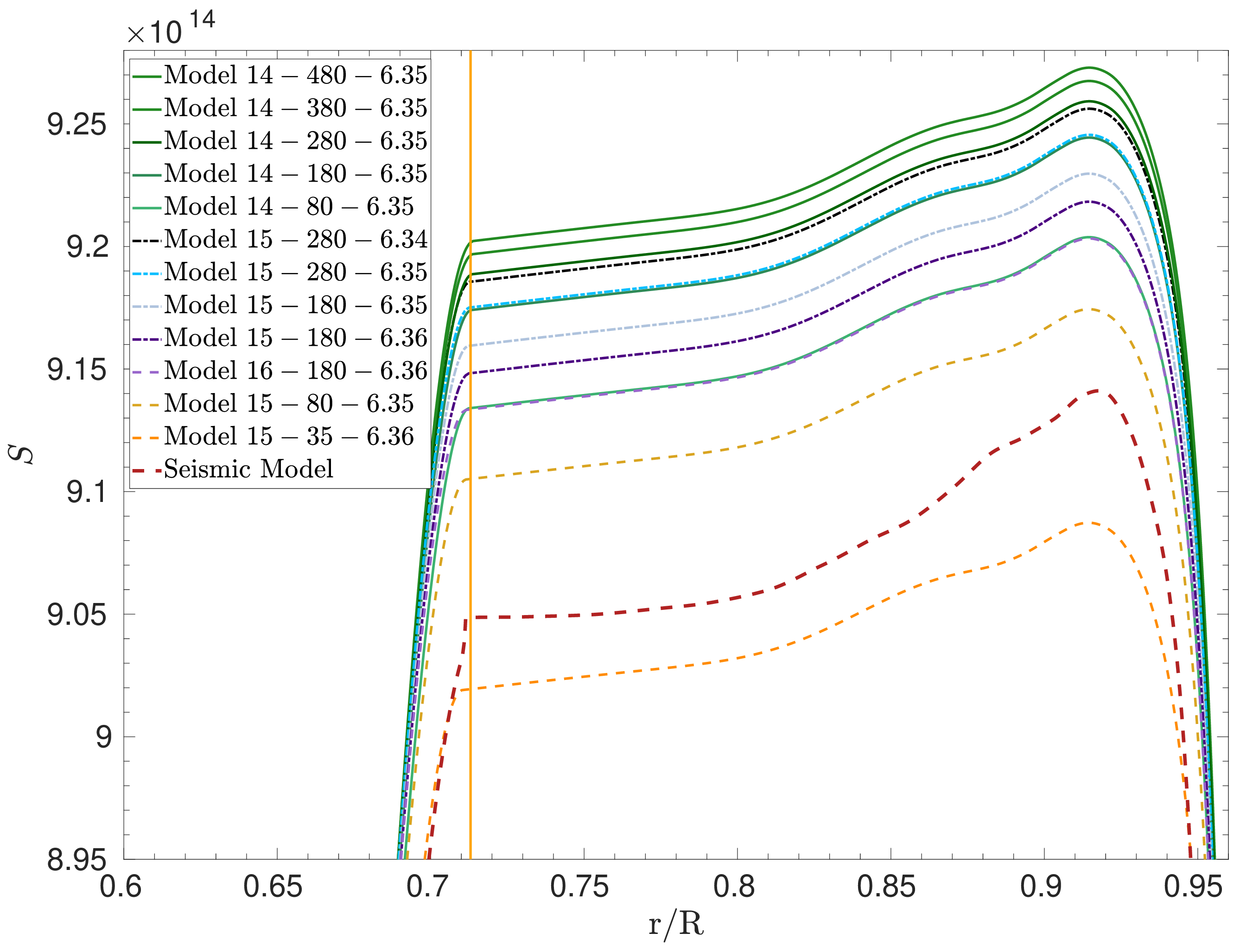}
	\caption{The entropy proxy profile as a function of normalized radius at the age of the Sun for the models of Table \ref{tabModelsOPAC}. The entropy proxy profile of a seismic model is also provided for comparison.}
		\label{Fig:SOpac}
\end{figure}

We present in Fig. \ref{Fig:SOpac} the impact of the opacity modification on the height of the entropy plateau. The first conclusion we can draw from looking at Fig. \ref{Fig:SOpac} and Table \ref{tabModelsOPAC} is that the opacity modification allows to alter very efficiently the height of the entropy plateau without changing much global parameters such as $\left(r/R\right)_{\rm{BCZ}}$, $\mathit{\rho}_{\rm{BCZ}}$ or even $\mathit{T}_{\rm{BCZ}}$. This is in stark contrast with the effects of adiabatic overshooting that significantly alter the BCZ properties and require unreasonable values to actually lead to a low enough entropy plateau. Looking at the effect of opacity and equation of state illustrated in Fig. \ref{Fig:SModMix}, we can see that these are much more promising avenues to reproduce the entropy proxy plateau in solar models. 

\begin{figure*}
	\centering
		\includegraphics[width=16cm]{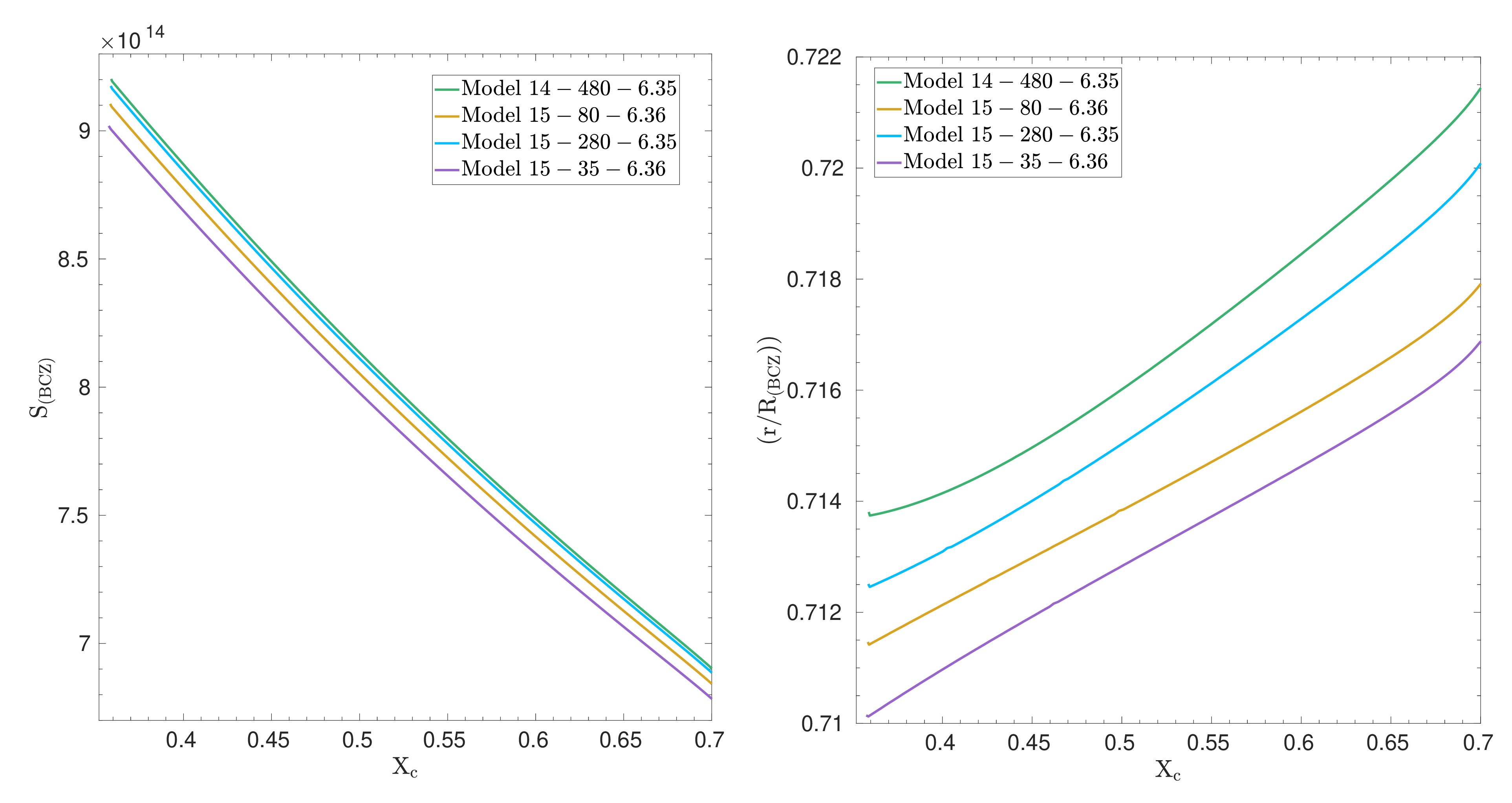}
	\caption{Evolution of the height of the entropy proxy plateau at the BCZ and of the depth of the BCZ as a function of central hydrogen mass fraction for three representative models of Table \ref{tabModelsOPAC}. Left panel: evolution of the height of the entropy proxy plateau, S(cz). Right panel: evolution of the position of the BCZ.}
		\label{Fig:SOpacEvol}
\end{figure*}

Unlike the effects of overshooting, there is also no correlation between the depth of the convective zone and the height of the entropy plateau as the model evolves. This is confirmed by looking at Fig. \ref{Fig:SOpacEvol} which plots in parallel the evolution of the height of the entropy plateau at the BCZ and the position of the BCZ as a function of central hydrogen. Again the evolution is very smooth, with the height of the plateau evolving regularly as a the model burns hydrogen. Again the plateau rises over time as the BCZ position evolves inwards. 

From a more detailed investigation of the height of the plateau and its link with the opacity increase, we can see that either increasing the amplitude, $A$, the width, $\sigma$ or the reference temperature at which the increase occurs, $\log T_{\rm{ref}}$ allows to push the entropy plateau down. It seems that increasing $A$ is also directly linked with a lower BCZ position, whereas the width of the opacity modification has a more efficient impact on the height of the plateau. 

However, these effects are clearly nonlinear and correlated; for example, changing the reference temperature of a narrow opacity increase only has a minor impact whereas changing only by $1\%$ the amplitude of a broader opacity peak can shift the plateau significantly downwards. The situation is thus much more complex than for adiabatic overshooting and it would not be really useful to attempt to parametrize exactly the required functional shape of the opacity modification that would lead to the correct height of the entropy plateau at the age of the Sun. 

\subsection{Origins of the variations of the entropy proxy plateau}\label{Sec:DominantFact}

The results presented in Sects. \ref{Sec:Overshooting} and \ref{Sec:Opacity} show that the entropy proxy plateau may be influenced by both adiabatic overshooting and an opacity increase. However, adiabatic overshooting leads to solutions in full disagreement with helioseismic constraints and it seems impossible to induce a readjustement of the entropy plateau as low as the value found by seismic models on the sole basis of an extension of the adiabatic layers of the deep convective envelope. The situation is the opposite for the opacity increase, for which large readjustments of the height of the entropy plateau can be obtained without being in conflict with the position of the BCZ. Using the definition of the entropy proxy and assuming the perfect gas equation, which is valid at the base of the convective zone, we can identify the contributing factors to its height 
\begin{align}
S=\frac{P}{\rho^{\Gamma_{1}}}=\frac{\mathcal{R}T}{\mu \rho^{\Gamma_{1}-1}}, \label{eq:SPGaz}
\end{align}

The first factor is temperature, with a lower temperature leading to a lower entropy plateau. This is quite difficult to achieve as the temperature of the BCZ is directly linked with the depth of the BCZ. As can be seen from Table \ref{tabModelsOPAC}, a given $(r/R)_{\rm{BCZ}}$ leads almost exactly to a given $T_{\rm{BCZ}}$ (and a given $(m/M)_{\rm{BCZ}}$). Reducing the temperature at the BCZ would require to push it outwards and go against the precise helioseismic determination of its location. Another option is to increase the mean molecular weight in the CZ. This quantity is, however, directly fixed by spectroscopic and helioseismic constraints. There is therefore little room for manoeuver for a given set of constraints, especially if one considers light element depletion \citep{Buldgen2025Be}.

The mean molecular weight also has a more intricate effect. For example, one might argue that a standard solar model using a high metallicity value produces a better entropy proxy plateau height than a model using the AAG21 abundances, as can be seen from Fig. \ref{Fig:SvsS53}, and that therefore this property validates a high metallicity in the solar envelope. This is incorrect as for example the AAG21 model including macroscopic mixing has almost the same mean molecular weight in the CZ as a GN93 model, as it has almost the same helium mass fraction in the CZ, but its entropy proxy plateau is significantly worse. This is due to its shallower convective zone that leads to a significantly lower density at the BCZ and thus a lower plateau. The better performance of a high metallicity SSM is not a result of the mean molecular weight value, but rather of the intrinsically larger opacity close to the BCZ that results from the higher metallicity. Adding mixing to reproduce the lithium depletion in such a model would lead to a worsening of the agreement of the entropy proxy plateau, while a significant variation is still required to actually agree with the seismically determined value. It would therefore be incorrect to use an entropy proxy inversion to validate the abundances, similarly to using a sound speed inversion to that end. 

The influence of the mean molecular weight implies, however, that changes may be observed if the conditions of a solar evolutionary model at the ZAMS are altered, such as the efficiency of nuclear reactions, the opacity at high temperature or the effects of planetary accretion. Such effects would need to be investigated in detail in future. 

\begin{figure*}
	\centering
		\includegraphics[width=15cm]{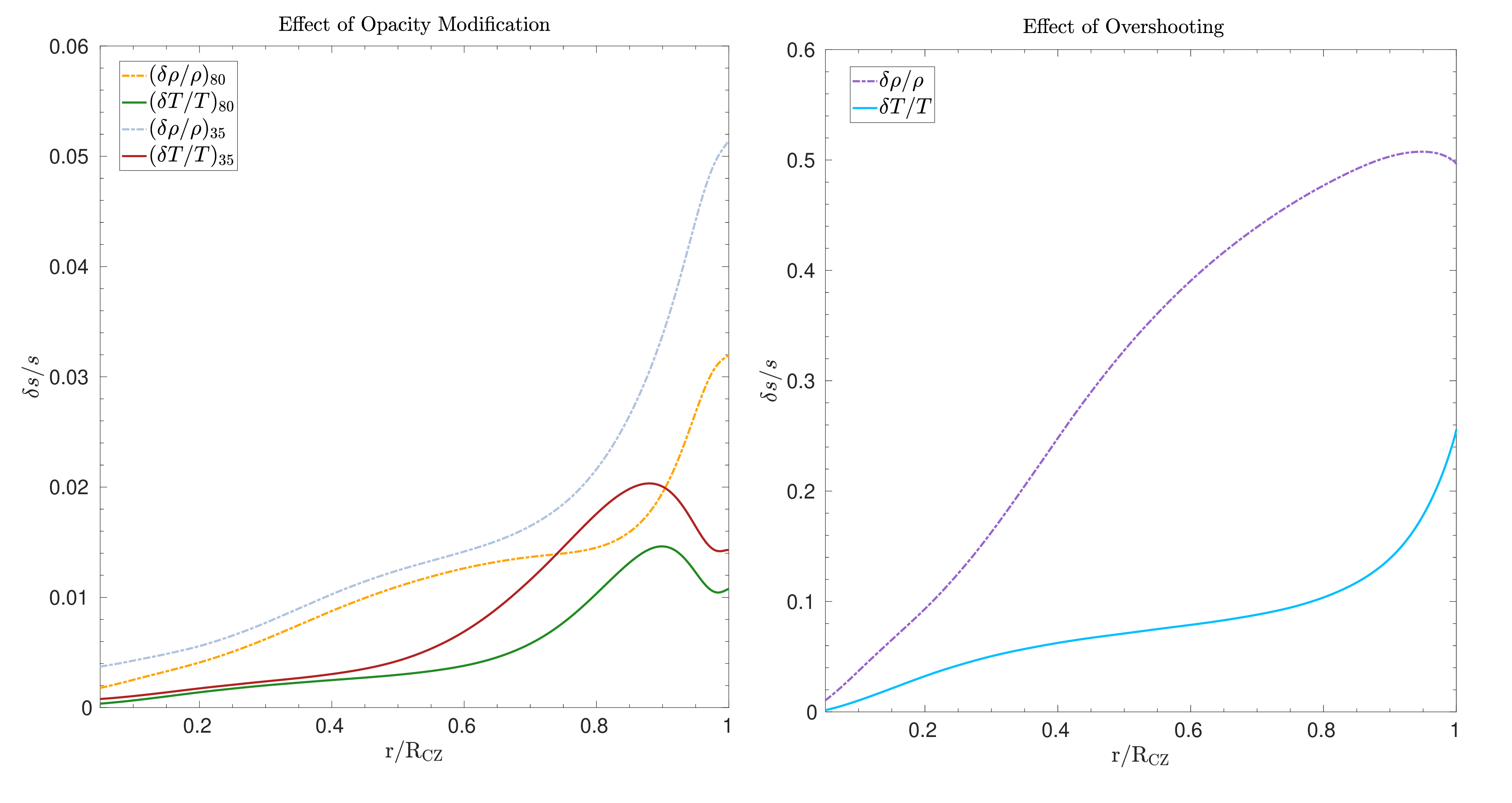}
		\caption{Relative differences in density and temperature as a function of the radius normalized at the position of the BCZ. Left panel: Effect of an opacity modification, as seen between Model $14-480-6.35$, Model $15-35-6.36$ and Model $15-80-6.35$. Right panel: Effect of overshooting, as seen between Model $0.65\rm{H_{P}}$ and Model $0.1\rm{H_{P}}$ .}
		\label{Fig:TRho}
\end{figure*}

The last quantity in Eq. \ref{eq:SPGaz} that may impact the entropy proxy plateau is the density in the CZ. This is again directly seen in Table \ref{tabModelsOPAC}: the model with the lowest entropy plateau in Fig. \ref{Fig:SOpac} also has the highest density at the CZ. The same stays true for the models including overshooting in Table \ref{tabModelsOv}, but in the case of overshooting, the higher density is accompanied by a significantly higher temperature (since the CZ is pushed inwards) which reduces the efficiency with which the entropy proxy plateau can be lowered. To understand what is going on, it is interesting to look at the temperature and density variations between the various models. In Fig. \ref{Fig:TRho}, we plot the relative differences in temperature and density between Model $0.1\rm{H_{P}}$ and Model $0.65\rm{H_{P}}$ for the overshooting models and Model $14-480-6.35$ and both Model $15-80-6.36$ and Model $15-35-6.36$ for the modified opacity models as a function of the radius normalized at the BCZ. This means that the upper limit of the plot is the BCZ and that the differences are renormalized so that the extent of the plateau is the same in both models. This choice is made so that the comparisons are made in the same relative distance with respect to the convective envelope. 

From Fig. \ref{Fig:TRho}, we can confirm that the differences in temperature and density are larger by an order of magnitude in the case of overhooting. The increase in density is of 49$\%$ a value that in itself would be sufficient to drastically reduce the height of the entropy proxy plateau. It is, however, accompanied by an increase in temperature by about 26$\%$, which, as seen in Eq. \ref{eq:SPGaz}, compensates for the density increase. All in all, if we estimate the change in the value of $S$ from these changes, we should see a decrease between Model $0.65\rm{H_{P}}$ and Model $0.1\rm{H_{P}}$ for the value of $S$ at the BCZ of about $3\%$, which is exactly what is seen from Fig.~\ref{Fig:SOver}.

If we look at the differences induced by an opacity modification, we can see that the density is changed by about $3\%$, while the temperature only changes by $1\%$. Again using Eq. \ref{eq:SPGaz}, the induced modification on the entropy proxy plateau is of about $1\%$, which is confirmed when looking at Fig. \ref{Fig:SOpac}. The same effect is observed for Model $15-35-6.36$, with the temperature and density modifications having essentially the same shape as for Model $15-80-6.36$, but a wider extension and a larger amplitude. This confirms that the increase in density at the BCZ required to lower the height of the entropy proxy plateau must be generated by a physical mechanism that does not affect temperature in a signicant way. In a more quantitative way, if $\delta \rho / \rho$ is the relative change in density, we simply must have $\delta T / T < (\Gamma_{1}-1)\delta \rho / \rho$. While this is the case for both physical processes here satisfy this rule, overshooting is accompanied by two main drawbacks. First, it must keep a sharp temperature gradient transition located at $0.713\rm{R}_{\odot}$. Second, it must not induce a too high depletion of lithium at the solar age. Model $0.65\rm{H_{P}}$ is in total contradiction with both these constraints and, given the large variations in density, likely does not lead to an adequate reproduction of the solar sound speed profile, while an opacity modification has been shown on numerous occasions to improve the agreement of models with helioseismic data. This again confirms that overshooting alone cannot be invoked to reproduce the height of the entropy proxy plateau in the Sun. 

\section{Additional thermodynamic quantities}\label{Sec:ThermoQuantities}

\begin{figure*}
	\centering
		\includegraphics[width=16cm]{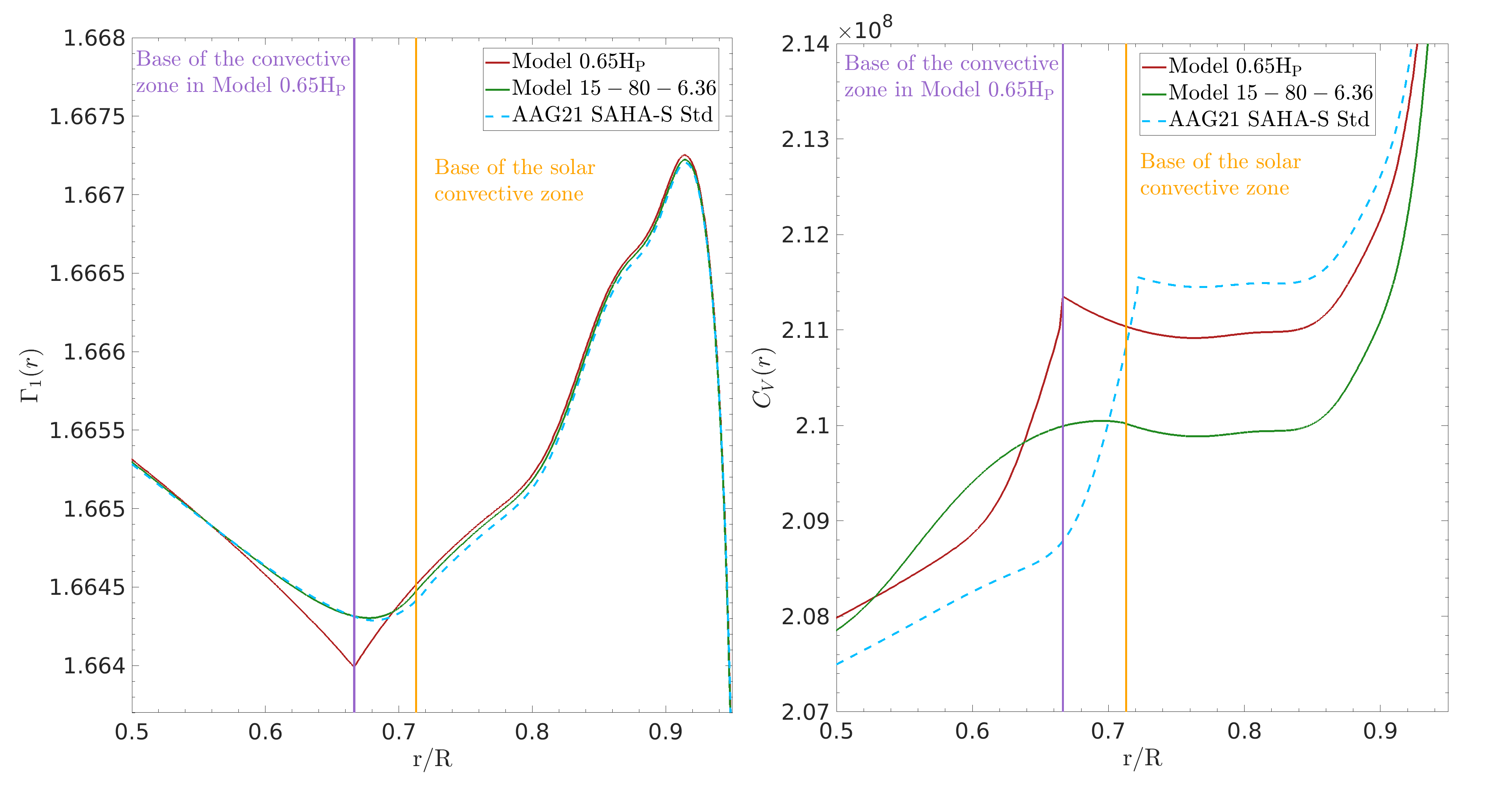}
		\caption{Left panel: First adiabatic exponent, $\Gamma_{1}$, as a function of normalized radius for an SSM, Model $0.65\rm{H_{P}}$ and Model $15-80-6.35$.; Right panel: Specific heat at constant volume $C_{V}$ as a function of normalized radius for an SSM, Model $0.65\rm{H_{P}}$ and Model $15-80-6.35$.}
		\label{Fig:G1Model}
\end{figure*}

In addition to looking at the main driving factors of the height of the entropy proxy plateau, it is also interesting to look at the behaviour of other thermodynamic quantities that are directly impacted by these model changes. In Fig. \ref{Fig:G1Model}, we illustrate in the left panel the $\Gamma_{1}$ profile of the standard solar model, Model $15-80-6.35$ and   Model $0.65\rm{H_{P}}$ and in the right panel the $C_{V}$ profile for the same models. Again, the profiles for $C_{V}$ are quite different, especially in the convective envelope, as a result of the difference $X_{CZ}$ in the value of $C_{V}$ in the convective envelope. Indeed, we added macroscopic mixing in the models including an opacity modification, which leads to a lower value of $X_{CZ}$ compared to the models with overshooting, as can be seen from Tables \ref{tabModelsOv} and \ref{tabModelsOPAC}. The higher $X_{CZ}$ value in the models including macroscopic transport leads to a much lower value of $C_{V}$ in the convective zone.

\begin{figure}
	\centering
		\includegraphics[width=8cm]{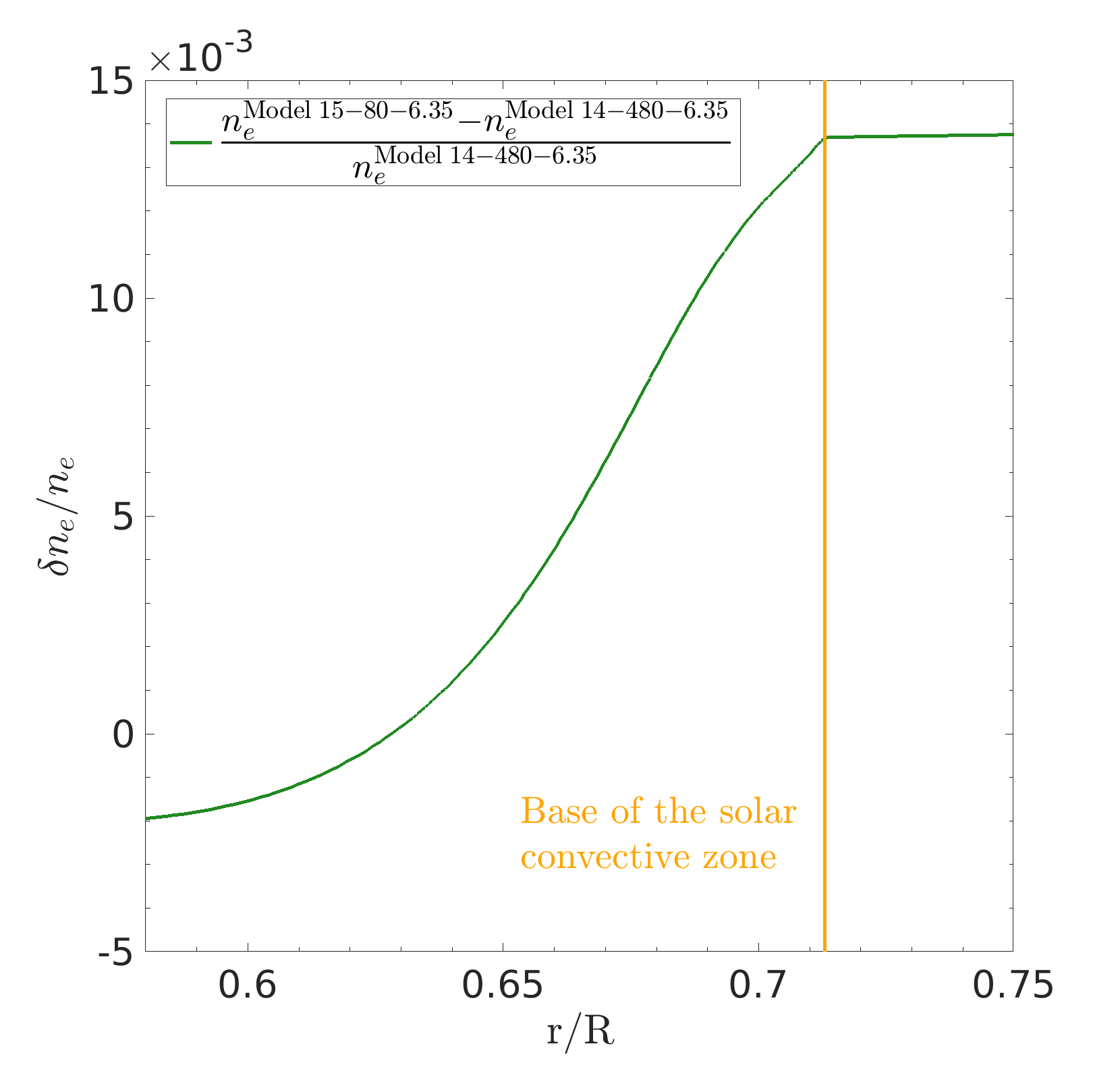}
		\caption{Relative differences in electronic density between Model $14-480-6.35$ and Model $15-80-6.35$ as a function of normalized radius.}
		\label{Fig:YeModel}
\end{figure}

The variations in $\Gamma_{1}$ induced by the change in opacity at the BCZ are very small, which implies a relatively small variation of the ion fractions at the BCZ, whereas the change induced by overshooting is much more extreme and leads to a clear dip in $\Gamma_{1}$ that would create a very clear signature. The small modification of ion populations is confirmed when looking at the relative differences in electronic density, $\rm{n_{e}}$, between Model $14-480-6.35$ and Model $15-80-6.35$, illustrated in Fig. \ref{Fig:YeModel}. We can note an increase of about $1\%$ of the electronic density close to the CZ that follows directly the change in density. 

Such a variation remains small, but is still significant at the level of precision required by helioseismic constraints. Moreover, it is to be noted that the evolution during the main sequence of the electronic density at the BCZ is far larger than the evolution of the temperature of the BCZ over the same period. Therefore, it might explain why a parametrization of opacity based on temperature solely will not capture the full evolutionary trends at the BCZ. On the other hand, adding a dependency on electronic density in an opacity correction requires a better understanding of its overall effect and ideally a guess of the physical origins of such a dependency.

\section{Conclusion}\label{Sec:Conc}

In this study, we have introduced an updated formulation of the entropy proxy indicator defined in \citet{Buldgen2017S}. We started in Sect. \ref{Sec:SPlateau} by introducing the definition of the indicator, its link with entropy and the differential equations to derive the structural kernels to carry out helioseismic inversions. We then study in detail in Sect. \ref{Sec:BCZ} the diagnostic potential of the indicator and its link to the two main modelling ingredients influencing its height in the convective envelope, namely convective overshooting at the BCZ and an opacity increase at the BCZ. We also discuss the link between the behaviour of this entropy plateau during the main sequence and key quantities such as the position, density and temperature at the BCZ. We find that we can strongly differentiate between the effects of overshooting and an opacity increase at the BCZ, concluding that the former cannot be invoked to solve the solar modelling problem linked with the revised abundances of \citet{Asplund2021}. The behaviour of the entropy proxy plateau in the CZ, coupled with the precise determination of the position of the BCZ in the Sun leaves an opacity increase to be the prime candidate to reduce the observed discrepancies. The exact magnitude and extent of that increase will be impacted by other physical ingredients such as the equation of state or the nuclear reaction rates that may alter the calibrated solar parameters such as its initial hydrogen mass or heavy element mass fraction. Our tests using the currently available physical inputs (see Fig. \ref{Fig:SModMix}) show that at least the equation of state and the modelling of convection available in CLES have little impact, but this remains only valid for the tested ingredients\footnote{A similar effect can be expected for changing the nuclear reaction rates within the tests carried out in \citep{Buldgen2019}.}. In this context, further work is needed to investigate the development of a comprehensive framework including the height of the entropy plateau in an extended calibration scheme. This however requires more detailed investigations.

In Sect. \ref{Sec:DominantFact}, we have investigated the dominant factors that may influence the behaviour of the entropy proxy plateau. We highlighted the fact that, for a given mean molecular weight, the density and temperature modifications must obey a simple proportionality law to allow for the height of the plateau to be lowered. This puts constraints on the physical mecanism that may allow to lower the entropy plateau in solar evolutionary models and confirms that overshooting is not the prime candidate to correct solar evolutionary models, as the modifications induced in the radiative zone are extreme. 

In Sect. \ref{Sec:ThermoQuantities}, we have looked into more detail at other thermodynamical quantities such as the specific heat, the first adiabatic exponent and the electronic density. We find that opacity modifications lead to small variations of electronic density that closely follow the density modifications. The changes induced in the specific heat profile and $\Gamma_{1}$ profile indicate that the overall modifications induced by the opacity modifications remain overall quite small, while still significant. A better understanding of the impact of such changes and their response to revision of physical processes entering the computation of opacities might be key to explaining the current discrepancies between theoretical opacity computations, experimental measurements and helioseismic determinations of the solar radiative opacity \citep{Buldgen2025b}.

Our investigation highlights the diagnostic potential of the entropy proxy indicator to unravel the evolution of the BCZ condition on the main sequence. Coupled to a detailed chemical evolution based on light element depletion \citep[following e.g.][]{Richard1996, Richard1997, Brun2002, Buldgen2025Be}, further analyses of the behaviour of specific heat at the BCZ and of the evolution of the entropy proxy plateau may help us further understand the required behaviour and dominant factors in the required opacity modifications. So far, we have only investigated the effects of modifications parametrized on temperature, while electronic density also plays a key role in influencing opacity. In this respect, the fact that the entropy proxy plateau might act as a key constraint to understand the shape of opacity corrections that would allow to replace its height at the seismic value and potentially help pinpointing missing physical processes in current opacity computations.

Evolutionary computations by \citet{Kunitomo2021} and \citet{Kunitomo2022} have already shown that a wider Gaussian provided a good agreement in sound speed. As demonstrated here, the entropy proxy plateau provides an independent diagnostic that allows to further characterize the shape, position and amplitude of an opacity modification. By linking it to microphysical ingredients of solar models at the BCZ, this diagnostics allows to efficiently complement Ledoux discriminant inversions \citep{Buldgen2017A} that have been used to infer the required opacity modification at the current solar age \citep{Buldgen2025b}. With this additional helioseismic diagnostic coupled to revised chemical transport prescription anchored in recent observations \citep{Wang2021, Amarsi2024}, we may provide an additional helioseismic diagnostic to complement experimental measurements that are pointing to further inaccuracies between opacity computations \citep{Bailey,Nagayama2019,Mayes2025}.

\begin{acknowledgements}
We thank the referee for their careful reading of the manuscript. GB acknowledges fundings from the Fonds National de la Recherche Scientifique (FNRS) as a postdoctoral researcher. The study by V.A.B., A.V.O., S.V.A. was conducted under the state assignment of Lomonosov Moscow State University.
\end{acknowledgements}

\bibliography{biblioarticleUnder}

\appendix

\section{Additional figures}\label{sec:AppendixA}
\begin{figure}
	\centering
		\includegraphics[width=8cm]{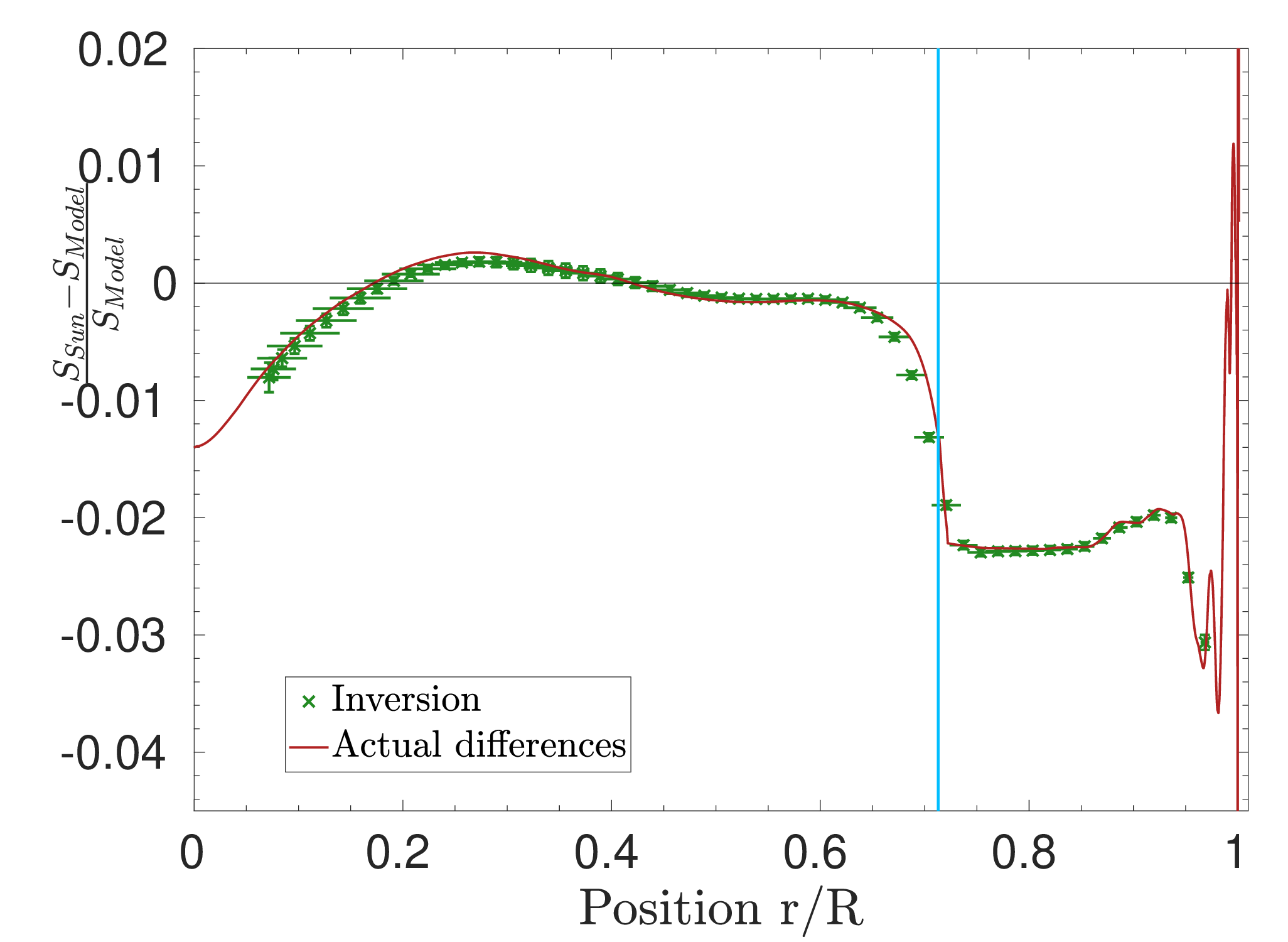}
         \caption{Example of inversion for the entropy proxy $S$ using artificial data, using an AAG21 SSM model as reference and a GN93 SSM as target.}
		\label{Fig:SHH}
\end{figure}

We illustrate in Fig. \ref{Fig:SHH} an inversion exercise on artificial data for the entropy proxy indicator $S$. The inversion accurately reproduces the height of the plateau and the overall features of the profile, while some degree of resolution limitation can be seen in the tachocline region. It is likely that non-linear RLS methods \citep{Corbard1999} will be required to fully resolve the profile in these layers. 

\end{document}